\documentclass[aps,prl,showpacs,reprint,superscriptaddress,footinbib]{revtex4-1}

\usepackage{amsmath}
\usepackage{graphicx}
\usepackage{dcolumn}
\usepackage{bm}

\usepackage{hyperref}

\hypersetup{
    pdffitwindow=true,     
    pdfstartview={FitH},    
    colorlinks=true,       
    linkcolor=blue,          
    citecolor=blue,        
    filecolor=magenta,      
    urlcolor=blue,           
    bookmarksdepth=-2,   
    pdfpagemode=UseNone   
}

\begin{document}

\title{Observation of an in-plane vortex lattice transition in the multiband superconductor 2$H$-NbSe$_{2}$ using scanning tunneling spectroscopy} 

\author{I. Fridman}
\affiliation{Department of Physics, University of Toronto, 60 St. George St., Toronto ON M5S1A7 Canada}

\author{C. Kloc}
\affiliation{School of Materials Science and Engineering, Nanyang Technological University 639798 Singapore}

\author{C. Petrovic}
\affiliation{Condensed Matter Physics and Materials Science Department, Brookhaven National Laboratory, Upton, NY 11973 USA}
\affiliation{Canadian Institute for Advanced Research, Toronto, ON, M5G1Z8 Canada}

\author{J. Y.T. Wei}
\affiliation{Department of Physics, University of Toronto, 60 St. George St., Toronto ON  M5S1A7 Canada}
\affiliation{Canadian Institute for Advanced Research, Toronto, ON, M5G1Z8 Canada}

\begin{abstract}
We report real-space evidence for an in-plane vortex lattice (VL) transition in the multiband superconductor $2H$-NbSe$_2$, observed by laterally imaging subsurface vortices with scanning tunneling spectroscopy. With magnetic field applied $\parallel$ $ab$-surface up to 6.5 T at 300 mK, spatial maps of the zero-bias conductance revealed a stripe pattern whose dominant periodicity showed a discrete shift near 0.7 T.  The stripes can be interpreted as the surface projection of a distorted hexagonal VL that undergoes a first-order reorientation transition.  This transition occurs for field $\parallel$ [100] but not [110], and is correlated with the multigap characteristics seen in the heat capacity and thermal transport data.  These results implicate the multiband pairing of $2H$-NbSe$_2$ in the VL transition we observed.

\end{abstract}


\maketitle

In the mixed state of type-II superconductors, a diamagnetic current flows along the sample edge and paramagnetic currents loop through an ordered array of flux quanta in the bulk to form a vortex lattice (VL) \cite{Abrikosov}.  The superconducting VL is known to be sensitive to crystal anisotropy, vortex-vortex interaction, pairing symmetry and band structure \cite{Brandt95,Kogan98,Machida10,Agterberg11}.  These factors can cause the VL to undergo geometric variations and thermodynamic transitions, which have been observed in various superconductors by both reciprocal-space and real-space measurements \cite{Eskildsen97,Renner95,Dewilde97,Yethiraj99,Kadono06,White09,Eskildsen12,Biswas12,Sakata00,Sosolik03}.  A transition of particular interest involves the rotation of a hexagonal VL vs. $c$-axis magnetic field, observed by neutron scattering in the multiband superconductor MgB$_2$ \cite{Cubitt03}.  This VL transition is second-order in character, and is correlated with the field-induced suppression of pairing in one the bands. In general, field depairing in a bulk superconductor is a function of its energy gap and Fermi velocity.  When the pairing involves multiple bands \cite{Suhl59, Schopohl77, Binning80} of different dimensionalities, as in the case of MgB$_2$, the depairing can thus proceed band by band and depend on the field direction \cite{Blouquet02, Machida04}. Given the discreteness and anisotropy of this depairing process, it is natural to consider whether other types of VL transitions could occur in such multiband superconductors.

In this Letter we report real-space evidence for a first-order VL transition in superconducting $2H$-NbSe$_2$, observed by scanning tunneling spectroscopy (STS) with magnetic field $H$ applied parallel to the $ab$-plane.  $2H$-NbSe$_2$ is a well-studied superconductor that has shown abundant evidence for multiband pairing \cite{Boaknin03, Sologubenko03,Rodrigo04, Sonier05, Johannes06, Kiss07, Huang07, Guillamon08, Noat10, Zehetmayer10, Rossnagel12}.  Spatial maps of the zero-bias tunneling conductance over the $ab$-surface, measured at 300 mK and up to 6.5 T, revealed a field-dependent stripe pattern that can be identified as the lateral projection of a distorted hexagonal VL.  While the stripe periodicity varies expectedly as $1/\sqrt{H}$, adjacent stripes show an amplitude alternation that increases with the field.  Fourier analysis of the pattern revealed two peaks whose relative intensities are reversed near 0.7 T, corresponding to a discrete shift in the dominant stripe periodicity.  This shift is observed for $H\parallel$ [100] but not [110], and is correlated with the multigap characteristics seen in field-dependent heat capacity \cite{Huang07} and thermal transport \cite{Boaknin03, Sologubenko03}.  We interpret this periodicity shift in terms of a VL reorientation transition that is driven by field depairing over parts of a multi-sheeted and anisotropic Fermi surface \cite{Johannes06, Fletcher07, Kiss07, Borisenko09}.  These results implicate the multiband pairing of $2H$-NbSe$_2$ in the VL transition we observed.

Our STS measurements were made on single crystals of 2$H$-NbSe$_2$, using a horizontalized scanning tunneling microscope housed inside a $^3$He refrigerator, with the field applied vertically by a superconducting solenoid.  The crystals were grown by an iodine-vapor technique \cite{Oglesby94}, structurally verified by x-ray diffraction (XRD), and showed superconducting critical temperatures $\approx$ 7.2 K.  The crystals were $\sim$ 5 x 5 x 0.5 mm$^3$ platelets with wide $c$-axis faces, and were aligned in-plane by XRD before being cleaved and cooled in zero field to 300 mK.  Topographic images showed atomically-resolved flat surfaces  (see Fig. S1 of the Supplemental Material \cite{supp}).  The $dI/dV$ conductance spectra were acquired using lock-in amplification with 20 $\mu$V excitation at 505 Hz, and the typical junction impedance was 10 M$\Omega$ at 4mV bias. Before measurement, the Pt-Ir tips were field-emitted \emph{in situ} to ensure junction stability over long scans, and RF-filtering was used throughout the cryogenic wiring to attain maximum spectral resolution.  Reproducibility of our measurements was checked over different scan areas and on different crystals.

\begin{figure}[t]
\includegraphics {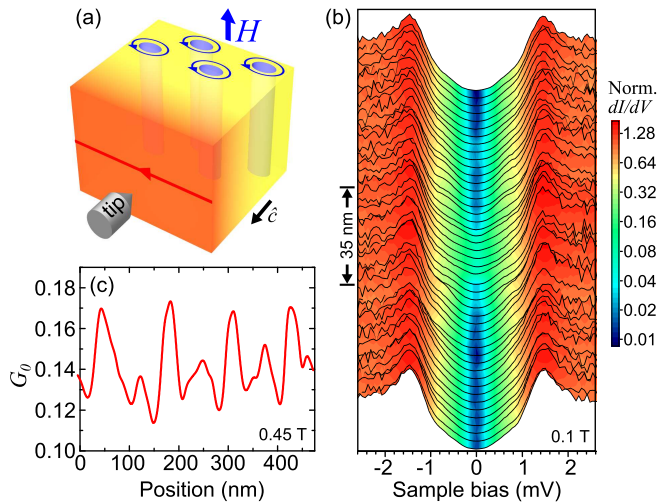}
\caption{\label{fig1}(color online). (a) Scanning tunneling spectroscopy on the $ab$-surface of $2H$-NbSe$_2$ at 300 mK, with a field-induced diamagnetic screening current (red line) flowing across the surface and paramagnetic currents (blue loops) encircling subsurface flux lines.  (b) Spatial variation of the normalized $dI/dV$ spectra, as the tip is scanned along a 140 nm line perpendicular to the in-plane field at 0.1 T.  A two-dimensional map of the normalized zero-bias conductance $G_0$ revealed a field-dependent stripe pattern as shown in Fig. \ref{fig2}a.  (c) Profile plot of the $G_0$ map across the stripes at 0.45 T, after line-averaging along the field, showing the alternating amplitude between adjacent stripes.}  
\end{figure}

\begin{figure*}
\includegraphics [width=17.8cm]{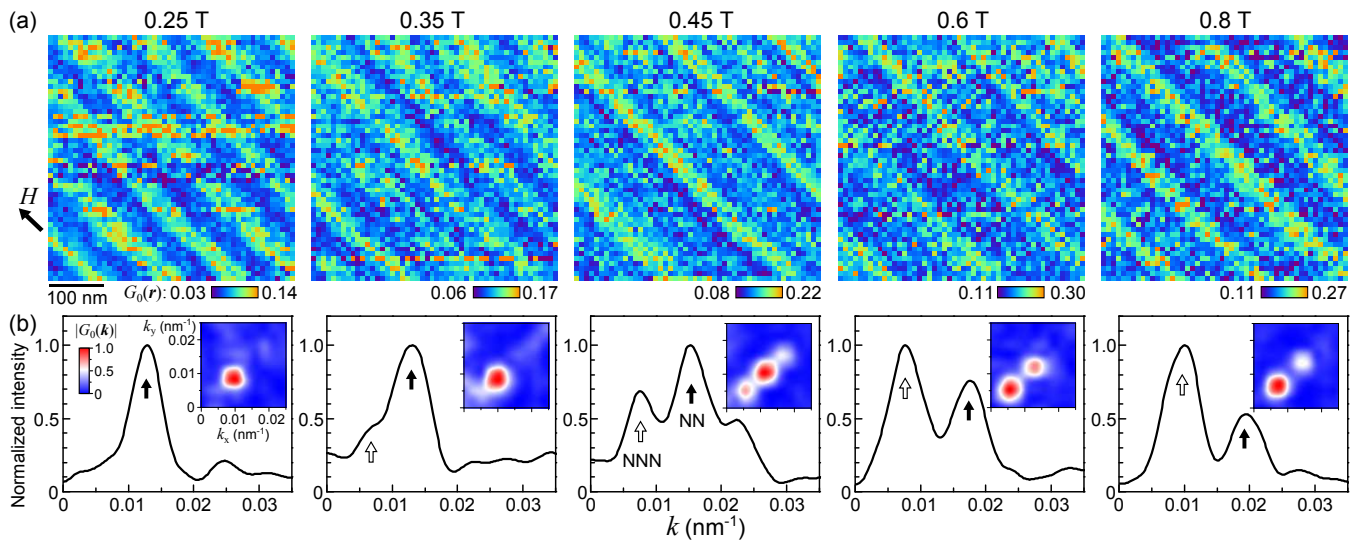}
\caption{\label{fig2}(color online).  (a) Spatial maps of the normalized zero-bias conductance $G_0(\bm{r})$, taken on the $ab$-surface of $2H$-NbSe$_2$ at 300 mK with field applied along [100] (black arrow).  Each 380 x 380 nm$^2$ map shows a pattern of stripes parallel to the field. The data for 0.25 T is reproduced from Ref. \onlinecite{FridmanAPL2011}.  (b) Fourier intensity $|G_0(\bm{k})|$ computed from each map, with the inset showing $|G_0(\bm{k})|$ in the positive quadrant of $k$-space, and the main plot showing a line trace of $|G_0(\bm{k})|$ from the origin through the Fourier peaks.  To facilitate comparison across fields, the intensity is normalized by the highest peak in each plot.  The alternating stripe amplitude gives rise to two Fourier peaks corresponding to the nearest-neighbor (NN, solid arrows) and next-nearest-neighbor (NNN, open arrows) stripe periodicities.  As the field is increased to 0.6 T and higher, the relative intensities of the NN and NNN peaks are reversed, consistent with fading of the minor stripes that effectively doubles the dominant stripe periodicity.}
\end{figure*}

Our cryomagnetic STS technique is illustrated Fig. \ref{fig1}.  The STS geometry is shown in Fig. \ref{fig1}(a), with a diamagnetic screening current sweeping across the sample surface, and paramagnetic vortex currents looping around flux lines in the bulk.  In contrast to previous STS studies which image cross sections of the VL \cite{Hess89,Hess92,Hess94, Kohen06, Suderow08}, our STS technique images the subsurface VL laterally and with depth sensitivity, by virtue of the interaction between surface and vortex currents \cite{FridmanAPL2011}.  Essentially, this interaction causes the field-dependent $dI/dV$ spectra to vary spatially, as shown in Fig. \ref{fig1}(b) for data taken on the $ab$-surface of 2$H$-NbSe$_2$ along a line perpendicular to the in-plane field at 0.1 T.  The spectra shown were normalized relative to the above-gap conductance at 3 mV.  A planar map of the similarly normalized zero-bias conductance $G_0(\bm{r})$, shown in Fig. \ref{fig2}(a), revealed a stripe pattern that is quantitatively related to the VL spacing.  Adjacent stripes alternate in amplitude, as shown in Fig. \ref{fig1}(c) by the stripe-crossing profile of $G_0(\bm{r})$ at 0.45 T, plotted after averaging along the stripes.  This amplitude alternation can be attributed to the depth alternation of the subsurface vortices, allowing us to associate the major (higher amplitude) and minor (lower amplitude) stripes with the first and second vortex rows, respectively.


\begin{figure}
\includegraphics {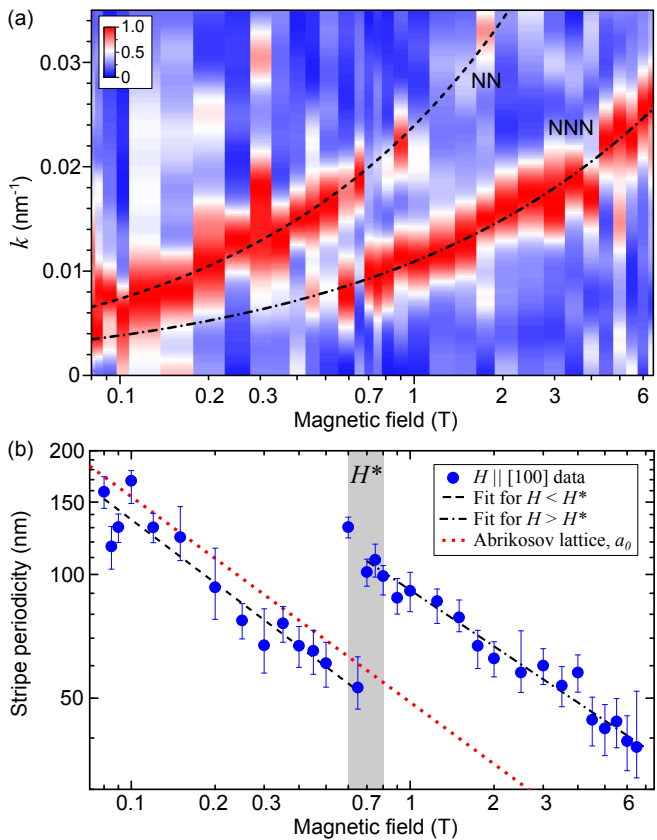}
\caption{\label{fig3}(color online).  Fourier analysis of the field evolution of $G_0(\bm{r})$.  (a) Line traces of $|G_0(\bm{k})|$ for fields between 0.08 T and 6.5 T applied along [100].  The two dashed curves are a guide to the eye, indicating square-root scaling of the peak positions.  The reversal of NN and NNN peak intensities near $H^*$= 0.7 T is associated with fading of the minor stripes and doubling of the dominant stripe periodicity.  (b) Log-log plot of the stripe periodicity corresponding to the dominant Fourier peak at each field.  The two dashed lines are power-law fits to the data below and above $H^*$, and the dotted line is the lattice parameter for a hexagonal Abrikosov lattice.  The sharp shift near $H^*$ shows no intermediate values of the stripe periodicity, consistent with a first-order transition.}
\end{figure}

\begin{figure}[ht!]
\includegraphics {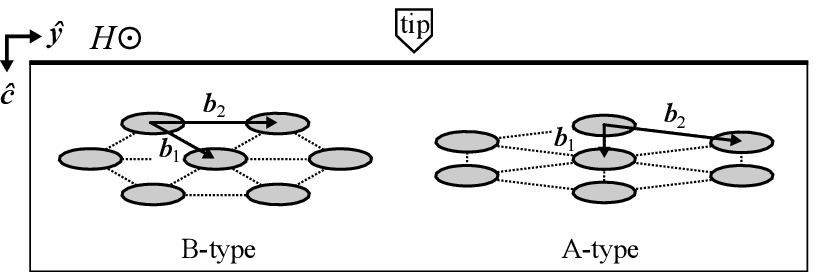}
\caption{\label{fig4} Two types of distorted hexagonal vortex lattices for anisotropic superconductors with the magnetic field applied in-plane.  The B-type and A-type lattices are sketched for the anisotropy parameter $\gamma$ being 3.0 and 4.8, respectively. Our STS technique essentially measures the lengths of the basis vectors $\bm{b}_1$ and $\bm{b}_2$ projected onto the sample surface.  Our data for $H\parallel$ [100] is consistent with a reorientation transition from B-type to A-type, which effectively doubles the predominant periodicity in the observed stripe pattern.}
\end{figure}


Identifying the stripe pattern as a lateral projection of the subsurface VL, we examine the field evolution of our $G_0$ data.  Fig. \ref{fig2}(a) displays a selection of $G_0(\bm{r})$ maps for $H\parallel$ [100] in the range of 0.25 to 0.8 T, which is well into the mixed state.  Each 380 x 380 nm$^2$ map shows a pattern of stripes parallel to $H$.  At 0.25 T, six bright stripes are seen, with no discernible alternation in amplitude. At higher $H$, the stripes increase in number and are largely unchanged in width, while amplitude alternation becomes noticeable.  As $H$ is increased to 0.6 T, the major stripes dominate over the minor stripes.  The minor stripes fade further at 0.8 T and are not discernible above 1 T, effectively doubling the stripe periodicity.

To quantify the stripe periodicities and amplitudes, we Fourier transform the $G_0(\bm{r})$ data and plot the normalized Fourier intensity $|G_0(\bm{k})|$ in Fig. \ref{fig2}(b).  In the insets, the $|G_0(\bm{k})|$ maps show prominent Fourier peaks evolving in both $k$-space location and intensity.  In the main panels, the intensities of these peaks are shown by line traces of $|G_0(\bm{k})|$ from the origin through the Fourier peaks.  Looking across the panels in Fig. \ref{fig2}, we associate the two prominent Fourier peaks with the nearest-neighbor (NN) and next-nearest-neighbor (NNN) stripe periodicities.  The NN peaks are approximately twice as far from the $k$-space origin as the NNN peaks, indicating that the major and minor stripes in $G_0(\bm{r})$ appear at regularly-spaced intervals.  At low fields, the NNN peak generally has lower intensity than the NN peak. However, as the field reaches 0.6 T and above, the relative peak intensities are reversed, consistent with fading of the minor stripes and doubling of the dominant stripe periodicity.

To visualize the reversal of NN and NNN peak intensities, we analyze how the Fourier peaks evolve over the entire field range of our measurement.  Fig. \ref{fig3}(a) shows a color plot of the line traces of $|G_0(\bm{k})|$, for fields between 0.08 T and 6.5 T applied along [100].  It can be seen that the NN peak dominates at low fields and the NNN peak at high fields, with a shift in their relative intensities occurring near 0.7 T.  The $k$ positions of both peaks appear to scale with $\sqrt{H}$, as indicated by the dotted curves.  We note that the intensity shift occurs only for $H\parallel$ [100] but not $H\parallel$ [110].  For the latter field orientation, the NN peak dominates the NNN peak both below and above $\sim$ 0.7 T.  This dependence on in-plane field direction is further discussed below, in the context of thermal transport data and band-structure and energy-gap anisotropies.

From the field evolution of the Fourier peaks, we obtain information about the in-plane VL.  Fig. \ref{fig3}(b) shows a log-log plot of the stripe periodicity versus field, which is the main result of this Letter.  The data points correspond to the dominant Fourier peak at each field, for $H\parallel$ [100].  It is clear that the data points fall along either of two lines, and show a discrete shift near 0.7 T, which we denote as $H^*$.  This periodicity shift is sharp, namely $\pm$0.1 T as defined by the reversal of the relative intensities of the Fourier peaks, and shows no intermediate values of the stripe periodicity.  By fitting the data points to the function $\beta H^\alpha$, we find that $\alpha$ is -0.51(5) for $H<H^*$ and -0.46(5) for $H>H^*$, both in good agreement with the expected 1/$\sqrt{H}$ dependence of the VL spacing.  On the other hand, we find that $\beta$ (in units of nm T$^{-\alpha}$) is respectively 42(3) and 92(6), indicating a scale change in the VL parameters.  It is convenient to express these results in terms of the Abrikosov VL parameter $a_0$ = $(4/3)^{1/4}\sqrt{\phi_0/B}$ for a hexagonal lattice, where $\phi_0$ is the flux quantum and $B$ is the magnetic induction.  For $H\parallel$ [100], the stripe periodicity is thus 0.86(7)$a_0$ below $H^*$ and 1.9(1)$a_0$ above $H^*$.

To interpret these stripe periodicities terms of a distorted VL, we consider the theory of anisotropic superconductors in the mixed state.  According to Ref. \onlinecite{Kogan88}, an in-plane field would produce a distorted hexagonal VL of either the A-type or B-type shown in Fig. \ref{fig4}, with the respective basis vectors:
\begin{align}
\bm{b}_1^{(A)}& = a_0\bm{\hat{c}}/\sqrt{\gamma} ,&
\bm{b}_2^{(A)}& = a_0(\bm{\hat{c}}/\sqrt{\gamma}+\sqrt{3\gamma}\bm{\hat{y}})/2 \nonumber\\
\bm{b}_2^{(B)}& = a_0\sqrt{\gamma}\bm{\hat{y}} ,&
\bm{b}_1^{(B)}& = a_0(\sqrt{3/\gamma}\bm{\hat{c}}+\sqrt{\gamma}\bm{\hat{y}})/2
\end{align}
where $\bm{\hat{y}}$ is parallel to the surface and perpendicular to the field, and $\gamma$ is an anisotropy parameter \footnote{Within the London framework of Ref. \onlinecite{Kogan88}, $\gamma = (m_{c}/m_{ab})^{1/2}$, where $m_{c}$ and $m_{ab}$ are the effective masses parallel and perpendicular to the the $c$-axis, respectively. This model has been extended to higher fields using anisotropic Ginzburg-Landau formalism \cite{Petzinger90}}. Surface projections of these vectors can be compared to the stripe periodicities we observed \footnote{For simplicity, we assume a hexagonal VL. We also assume that vortices near the surface are distributed with uniform density equal to that of the bulk \cite {Shmidt72, Brandt81}, although the distance between the surface and the first vortex row scales non-trivially with field \cite{Ternovskii72, Koshelev94}}.  Prior studies of $2H$-NbSe$_2$ have only seen the B-type VL: \textcite{Gammel94} observed $\gamma$ = 3.2(1) using neutron diffraction at 5.2 K and 0.8 T for $H\parallel$ [110]; \textcite{Hess94} observed $\gamma$ of up to 3.3 using STS at 300 mK and off $c$-axis fields up to 0.5 T.  Our STS data for $H\parallel$ [100] below $H^*$ and $H\parallel$ [110] are in good agreement with these prior results. That is, by associating our NN stripe periodicity with surface projections of the first two rows of vortices for the B-type VL, we get 0.86(7)$a_0$ = $\bm{b}_1^{(B)}\cdot\bm{\hat{y}}$ = $\bm{b}_2^{(B)}\cdot\bm{\hat{y}}$/2 with $\gamma$ = 3.0(5).  For $H\parallel$ [100] above $H^*$, a regime for which there have  been no prior measurements, our data cannot be easily reconciled with the B-type VL.  That is, in order to stretch $\bm{b}_1^{(B)}\cdot\bm{\hat{y}}$ from 0.86(7)$a_0$ to 1.9(1)$a_0$, $\gamma$ has to increase from $\approx$ 3 to 14, which is much larger than expected (see Supplemental Material \cite{supp} for a discussion of $\gamma$ in $2H$-NbSe$_2$).

To explain the large increase in stripe periodicity for $H\parallel$ [100] above $H^*$, we propose a reorientation transition of the B-type VL to the A-type VL.  In this scenario, a similar geometric analysis for the A-type VL, namely letting 1.9(1)$a_0$ = $\bm{b}_2^{(A)}\cdot\bm{\hat{y}}$, yields $\gamma$ = 4.8(5), which is more consistent with the available data on $2H$-NbSe$_2$ \cite{Gammel94,Hess94}.  For the A-type VL, it is important to note that the vortices along $\bm{b}_1^{(A)}$ are collinear with the tip and thus do not show up twice on the surface projection.  This geometric difference effectively doubles the periodicity between adjacent stripes as compared with the B-type VL. Thus, the reorientation transition between these two types of distorted hexagonal VL provides a natural explanation for the stripe periodicity shift. It should also be emphasize that the in-plane VL reorientation transition we observed is first-order in character, as evidenced by the discreteness of the periodicity shift. This observation is in contrast to the second-order VL transition observed in MgB$_2$ \cite{Cubitt03}, which involves the gradual rotation of an undistorted hexagonal VL for $H\parallel$ $c$-axis.

Our results can be directly correlated with the multigap characteristics seen in heat capacity \cite{Huang07} and thermal transport \cite{Sologubenko03} measurements on $2H$-NbSe$_2$.  First, the $H^*$ we observed coincides with the in-plane field at which both the Sommerfeld coefficient and thermal conductivity change slope.  These characteristics have been interpreted as a distinct signature of multigap pairing, similar to the case of MgB$_2$ \cite{Blouquet02}, whereby field-driven collapse of a weak gap as the field is increased above $H^*$ causes a decrease in the overall depairing rate.  Second, a hysteretic anomaly is seen in the thermal conductivity at 380 mK, occurring near $H^*$ for current perpendicular but not parallel to the in-plane field \cite{Sologubenko03}.  The hysteretic behavior is consistent with a first-order transition, and although the in-plane field direction was not specified in Ref. \onlinecite{Sologubenko03}, the directional dependence of this anomaly is remarkably similar to what was also seen in our data.  Thus the VL reorientation we observed provides a natural explanation for the unusual phase transition reported by Ref. \onlinecite{Sologubenko03}. These experimental correlations indicate a VL reorientation transition that is driven by field depairing over parts of a multi-sheeted Fermi surface having both uniaxial and in-plane anisotropies. This scenario is plausible considering the multi-sheeted and multi-gapped Fermi surface known to exist in $2H$-NbSe$_2$ \cite{Johannes06, Fletcher07, Kiss07, Borisenko09}.  Detailed theoretical study is needed to understand how the actual band structure and superconducting-gap anisotropy of $2H$-NbSe$_2$ affect its in-plane VL, following a similar analysis for MgB$_2$ \cite{Zhitomirsky04}.  

In the context of multiband pairing, we can also consider an alternative model for the stripe periodicity shift that involves a change in the vortex current distribution.  
Essentially, by regarding the vortex current as consisting of multiple superfluid components \cite{Tanaka09}, we can relate its radial distribution to the dimensionalities of the Fermi surface sheets that take part in the pairing.  For $2H$-NbSe$_2$ above $H^*$, field suppression of pairing on the 3D sheet \cite{Boaknin03, Huang07} would make the vortices less extended along $\bm{\hat{c}}$ and limit our ability to detect the second vortex row, thus doubling the dominant stripe periodicity.  The in-plane VL could then remain in the B-type orientation above $H^*$, without altering $\gamma$, and still produce the periodicity shift observed.  However, to validate this model, it would be necessary to calculate the vortex current distribution taking into account the actual band structure of $2H$-NbSe$_2$.  It would also be necessary to explain why the periodicity shift we observed occurs for $H\parallel$ [100] but not [110].

In summary, using cryomagnetic STS to laterally image subsurface vortices, we have observed a first-order transition of the in-plane VL in $2H$-NbSe$_2$ near 0.7 T at 300 mK. The transition is characterized by a discrete shift in the projected spacing of a distorted hexagonal VL, consistent with its field-driven reorientation. The field at which the transition occurs and its dependence on in-plane field direction are correlated with the multigap characteristics seen in both heat capacity and thermal transport. These results indicate that the multiband pairing of $2H$-NbSe$_2$ plays an important role in the VL transition we observed.

We acknowledge support by NSERC, CFI-OIT and CIFAR. Part of this work was carried out at BNL, which is operated for the U.S. Dept. of Energy by Brookhaven Science Associates (Grant No. DE-Ac02-98CH10886).

\end{document}


\title{Supplemental Material for \\ ``Observation of an in-plane vortex lattice transition in the multiband superconductor 2$H$-NbSe$_{2}$ using scanning tunneling spectroscopy''} 

\author{I. Fridman}
\affiliation{Department of Physics, University of Toronto, 60 St. George St., Toronto ON M5S1A7 Canada}

\author{C. Kloc}
\affiliation{School of Materials Science and Engineering, Nanyang Technological University 639798 Singapore}

\author{C. Petrovic}
\affiliation{Condensed Matter Physics and Materials Science Department, Brookhaven National Laboratory, Upton, NY 11973 USA}
\affiliation{Canadian Institute for Advanced Research, Toronto, ON, M5G1Z8 Canada}

\author{J. Y.T. Wei}
\affiliation{Department of Physics, University of Toronto, 60 St. George St., Toronto ON  M5S1A7 Canada}
\affiliation{Canadian Institute for Advanced Research, Toronto, ON, M5G1Z8 Canada}

\maketitle

\begin{flushleft} 
{\bf S1. Sample Topography}
\end{flushleft} 
\begin{figure}[h!]
\includegraphics [scale=1.1]{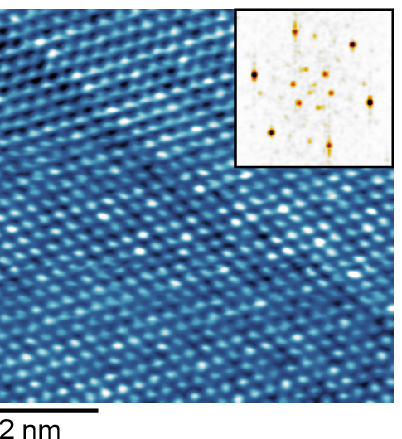}
\caption{\label{fig1} Atomically-resolved topographic image (7.5 x 7.5 nm$^2$) of $2H$-NbSe$_2$ at 300 mK showing hexagonally arranged Se atoms and charge-density-wave corrugation, both also visible in the Fourier transformed image (inset).}
\end{figure}

\begin{flushleft} 
{\bf S2. Anisotropy parameter of $2H$-NbSe$_2$}
\end{flushleft} 

In the context of a superconductor with uniaxial anisotropy, the anisotropy parameter $\gamma$ is defined as the ratio of superconducting length scales along the $c$-axis and in the $ab$-plane.  Early works considered $\gamma$ to be related to the effective mass anisotropy \cite{Kogan88}. In the case of multiband superconductors such as MgB$_2$, recent experiments \cite{Blouquet02,Cubitt03,Pribulova07} and theory \cite{Miranovic03} suggest that $\gamma$ is in general a function of the temperature $T$ and magnetic field amplitude $H$.  Since $2H$-NbSe$_2$ has considerable Fermi velocity ($v_F$) and superconducting gap ($\Delta$) anisotropy in the $ab$-plane, its $\gamma$ can also depend on the in-plane field orientation.  Here, we review previous measurements on $2H$-NbSe$_2$ and discuss the expected $\gamma$ for 300 mK and $H\parallel$ [100] above $H^*$, a regime that has not been previously studied.

$2H$-NbSe$_2$ is a superconductor with 3D, quasi-2D and 2D Fermi surface sheets \cite{Johannes06}.  At $T\ll T_c$, all the bands are involved in the pairing \cite{Boaknin03} and thus determine $\gamma$.  At $H>H^*$, however, $\gamma$ is expected to increase as pairing on the 3D sheet is suppressed, a phenomenon also seen in MgB$_2$ \cite{Bouquet02}.  $\gamma$ can be approximated as $\gamma_\lambda(T)$ = $\lambda_{\parallel c}$/$\lambda_{\perp c}$ near $H_{c1}$ and $\gamma_{H_{c2}}(T)$ = $H_{c2,\perp c}$/$H_{c2,\parallel c}$ near $H_{c2}$, where $\lambda$ is the London penetration depth, $H_{c1}$ and $H_{c2}$ are the lower and upper critical fields respectively, and the direction is denoted by the subscripts \cite{Miranovic03}.

Previously, \textcite{Hess94} measured $\gamma$ of up to 3.3, using scanning tunneling spectroscopy (STS) at 300 mK and off $c$-axis fields of up to 0.5 T, below $H^*$ = 0.7 T. Above $H^*$, due to field suppression of pairing on the 3D band \cite{Boaknin03, Huang07}, we expect a higher $\gamma$ than the one measured by \textcite{Hess94}. \textcite{Gammel94} measured $\gamma$ = 3.2(1) using neutron diffraction at 5.2 K and 0.8 T with $H\parallel$ [110].  However, since $T\sim T_c$ in the latter experiment, thermal mixing renders the system less anisotropic. Using the temperature dependence of $\gamma_{H_{c2}}$ \cite{Arai04} we extrapolate a $\sim$20$\%$ increase in $\gamma$ at 300 mK from the value of \textcite{Gammel94}.

We now discuss the dependence of $\gamma$ on in-plane field orientation.  At $H$ = 0.5 T, \textcite{Hess94} found a $\gamma$ that was $\sim$ 10$\%$ higher for $H\parallel$ [100] than for $H\parallel$ [110].  In light of recent measurements of the Fermi surface of $2H$-NbSe$_2$ \cite{Kiss07}, this can be understood as coming from the sixfold anisotropy of the 2D and quasi-2D bands in the $ab$-plane, which have $v_F$ minima and $\Delta$ maxima along $\Gamma-K$, or [110] in real space.  Qualitatively, this yields $\xi_{[100]}$ $>$ $\xi_{[110]}$, where $\xi$ = $\hbar v_F$/$\pi\Delta$ is the coherence length.  Thus, $H_{c2,[100]}$ $>$ $H_{c2,[110]}$ , where $H_{c2,[100]}$ = $\phi_0/2\pi\xi_{[001]}\xi_{[110]}$, $H_{c2,[110]}$ = $\phi_0/2\pi\xi_{[001]}\xi_{[100]}$ and $\phi_0$ is the flux quantum.  The anisotropy $\gamma_{H_{c2}}$ is then larger for $H\parallel$ [100] than for $H\parallel$ [110].  This effect is particularly significant at $H>H^*$ where the 2D and quasi-2D bands dominate.  

The results of the above analysis are consistent with the field dependence of $\gamma$ deduced from our data.  For $H\parallel$ [100] above $H^*$, we observed an A-type vortex lattice with $\gamma$ = 4.8(5).  For $H\parallel$ [110] both below and above $H^*$, we observed a B-type vortex lattice with a lower $\gamma$ of 3.0(5).  It is worth noting that a similar analysis has been performed for both ${H_{c2}}$ and $\gamma_{H_{c2}}$ in MgB$_2$ \cite{Zhitomirsky05}, but to our knowledge no such work has yet been carried out for $2H$-NbSe$_2$.  To further elucidate the effect of in-plane field orientation on the superconducting anisotropy of $2H$-NbSe$_2$, it would be vital to reexamine previous measurements of its $H_{c2}$ \cite{Arai04}, Sommerfeld coefficient \cite{Huang07} and thermal conductivity \cite{Sologubenko03} under known in-plane field orientation.